# Title: Synthetic Data & Health Privacy

## Subtitle: Should Generative AI Dream of Electric Patients?

**Authors**:

Gwénolé ABGRALL MD (1)(2)

Xavier MONNET MD, PhD (1)

Anmol ARORA MB, BChir (3,4)

(1)     AP-HP, Service de Médecine Intensive-Réanimation, Hôpital de Bicêtre, DMU 4 CORREVE, Inserm UMR S_999, FHU SEPSIS, CARMAS, Université Paris-Saclay, 78 rue du Général Leclerc, 94270, Le Kremlin-Bicêtre, France.

(2)     Service de Médecine Intensive Réanimation, Centre Hospitalier Universitaire Grenoble Alpes, Av. des Maquis du Grésivaudan, 38700, La Tronche, France.

(3)     School of Clinical Medicine, University of Cambridge, CB2 0SP, Cambridge, United Kingdom

(4)     University College London, Gower St, WC1E 6BT, London, United Kingdom

**Corresponding author's contact information:**

Dr Gwénolé ABGRALL

Service de médecine intensive-réanimation, Hôpital de Bicêtre, 78, rue du Général Leclerc, 94270 Le Kremlin-Bicêtre, France

+33650133671

gwenoleabgrall@gmail.com

**Wordcount**:  1190 words

While it remains to be seen whether Generative Artificial intelligence (GenAI) will genuinely revolutionise the core medical activity of clinical decision-making, it has already entered medical practice. AI-powered medical scribe tools (e.g., Nabla, Abridge, Ambience) have emerged as efforts to automate administrative tasks such as discharge summaries and referral notes, reducing physicians' workload. A recent survey revealed that many healthcare practitioners are now using publicly available large language models (LLMs)—a form of GenAI designed to process and generate text—for assisting in documentation, differential diagnosis, and exploring treatment options (1) .

GenAI creates new content, such as text or images, by learning from data patterns, (while "traditional" AI primarily focuses on classification and prediction). Effective healthcare applications of GenAI require vast amounts of high-quality, domain-specific data. However, as publicly available health data is exhausted, demand for private health data will increase, raising significant privacy concerns.

Synthetic data— designed to emulate real patient characteristics without revealing identifiable information—offers a potential solution to this conundrum. It can be generated using rule-based techniques, statistical modelling to approximate real data distributions, or by training GenAI models— such as generative adversarial networks (GANs)—on real data to capture underlying structures and create more complex synthetic datasets (2).

**Generative algorithms can sometimes "parrot" their training data**, a phenomenon known as memorisation, whereby models reproduce verbatim parts of their training corpus. This could lead to inadvertent leakage of copyrighted or, more concerningly, sensitive private information. This issue has already surfaced amongst popularly deployed GenAI models as it is central to current lawsuits between GenAI companies and creators or content owners (e.g., *Getty Images v. Stability AI*, *Authors Guild v. OpenAI*, *New York Times v. OpenAI*), where 'data leakage' is cited as key evidence of copyright violations.

More alarmingly, in late 2023, researchers exploited a flaw in ChatGPT 3.5. By prompting it to repeat words like "poem" indefinitely, the model eventually leaked sensitive training data, including phone numbers and email addresses, despite existing safeguards (3). Though it started by repeating "poem," it later diverged, unexpectedly recalling random training data extracts.

Notably, medical data may be especially vulnerable to memorisation due to its structured and repetitive nature, including standardised diagnostic codes or test results, simplifying patterns the model learns. Rare medical cases are even more prone to memorisation because models struggle to generalise from fewer examples, further increasing the risk of sensitive disclosures.

**The synthetic data generation market is booming**, projected to reach USD $2.89 billion with a compound annual growth rate of 60.02% from 2023 to 2028 (4). Sectors such as autonomous driving (e.g., Waymo), conversational AI (e.g., Amazon Alexa), and fraud detection (e.g., American Express) have already widely embraced synthetic data. In healthcare, applications include generating synthetic medical records and imaging. Use cases may include democratising access to data for researchers and to facilitate easier data sharing between institutions. Despite its

potential, however, synthetic data adoption in healthcare remains comparatively limited, due to the complexity and stringent regulation within the sector.

Open-access healthcare data is scarce and lacks geographic and demographic diversity (5). Privacy constraints further limit data scope, reducing its effectiveness for underrepresented populations and specific conditions, which are inherently associated with a higher risk of reidentification. These limitations risk producing biased models that reinforce healthcare disparities. Synthetic data could help by generating additional data to support fairer predictions, for example by selectively creating data for minority groups. However, **creating realistic and useful medical data remains challenging**, especially for rare conditions, dynamic changes, and outliers. In fields such as critical care, capturing intricate biological and clinical processes within highly detailed, multi-modal data is difficult, potentially reducing the accuracy and reliability of AI-generated insights. Synthetic data may also perpetuate or even amplify unresolved biases and spurious correlations from the original data, especially when produced at scale. Without careful evaluation and mitigation, this could exacerbate healthcare disparities rather than promote fairness and equity.

**Synthetic data is not inherently free from re-identification risks**.

When based on real patient data, synthetic data may overfit—replicating actual records—or be vulnerable to re-identification through techniques such as membership inference (detecting if a model used someone's data by analysing outputs) or linkage attacks (matching quasi-identifiers across datasets). Currently, no widely accepted standards exist for generating or evaluating synthetic healthcare data, and existing metrics alone remain insufficient. Evaluations should comprehensively assess fidelity, utility, and privacy, addressing inherent trade-offs among them.

**Nothing Personal, Or Is It?**

The AI Act, which came into force on 1 August 2024, promotes responsible AI development in the EU, introducing synthetic data as a privacy-preserving alternative to personal data in high-risk AI systems. Article 10 mentions synthetic data alongside anonymised data for processing sensitive information to detect and correct biases. Article 59 implies synthetic data as an alternative to personal data for AI development in regulatory sandboxes, using personal data only when synthetic or anonymised data cannot meet objectives. Whilst the recognition of synthetic data as a type of data is a positive step, the legislation does not discuss privacy standards for synthetic data to be considered non-personal data and instead automatically groups it with anonymised or other "non-personal" data.

Creating synthetic data from real personal data is considered processing under GDPR. However, whether and when synthetic data remains personal data—and thus subject to GDPR—remains a complex issue. Recent legislation points towards synthetic data as not being considered personal data. The traditional distinction between personal and non-personal data may not neatly apply to synthetic data, which varies in origins, methods, and purposes, often blurring this boundary (e.g., there is not always a one-to-one correspondence between the original and synthetic data). This orthodoxy is now being questioned by the proposal of a "privacy threshold" to assess re-identification risks (6). On 15 July 2024, Singapore's Personal Data Protection Commission Singapore (PDPC) released its *Proposed Guide on Synthetic Data Generation*. The Guide provides a practical, step-by-step framework for mitigating re-identification risks, managing residual risks, and ensuring data utility through ongoing monitoring, technical controls and governance measures.

**There is no silver bullet for managing privacy risks posed by GenAI, but combining strategies can help mitigate them**. Until synthetic data's legal status is clarified, healthcare stakeholders can take key precautions.

Clinicians and developers should avoid using sensitive data to train, fine-tune or employ GenAI models, reducing the risk of privacy breaches.

Open-source models can offer greater transparency in managing privacy risks compared to closed platforms, as seen with Nabla's recent shift from ChatGPT to open-source models, arguing for better control tailored to healthcare needs (7). Hosting LLMs on private cloud infrastructure can further reduce risks, though it does not fully eliminate the chance of sensitive data leakage. Similarly, combining Privacy Enhancing Technologies (PETs), such as, federated learning, synthetic data, and homomorphic encryption, can perhaps further strengthen data privacy while preserving utility.

Educating users and stakeholders about risks and proper applications of GenAI is crucial, as is establishing clear policies for handling and sharing generated content. While synthetic data presents a promising balance between innovation and privacy, challenges remain. Clear regulatory frameworks will be essential for unlocking its full potential, safeguarding patient safety and privacy, and adapting to rapid technological advancements.